\documentclass{PoS}

\newcommand{\hi}{{\sc H\,i}}
\newcommand{\mhi}{$M$(\hi)}
\newcommand{\msun}{{M$_\odot$}}
\newcommand{\kms}{$\,$km$\,$s$^{-1}$}

\title{The MeerKAT Fornax Survey}

\ShortTitle{The MeerKAT Fornax Survey}

\author{\speaker{P. Serra}$^{,1}$, W.~J.~G.~de Blok$^{2,3,4}$, G.~L.~Bryan$^{5,6}$, S.~Colafrancesco$^{7}$, R.-J.~Dettmar$^{8}$, B.~S.~Frank$^{4}$, F.~Govoni$^{1}$, G.~I.~G.~J\'{o}zsa$^{9,10,11}$, R.~C.~Kraan-Korteweg$^{4}$, S.~I.~Loubser$^{12}$, F.~M.~Maccagni$^{2,3}$, M.~Murgia$^{1}$, T.~A.~Oosterloo$^{2,3}$, R.~F.~Peletier$^{3}$, R.~Pizzo$^{2}$, M.~Ramatsoku$^{2,3,4}$, L.~Richter$^{9}$, M.~W.~L.~Smith$^{13}$, S.~C.~Trager$^{3}$, J.~H.~van~Gorkom$^{5}$, M.~A.~W.~Verheijen$^{3}$\\
        $^{1}$ INAF - Osservatorio Astronomico di Cagliari, Via della Scienza 5, 09047 Selargius (CA), Italy\\
        $^{2}$ ASTRON, Netherlands Institute for Radio Astronomy, Postbus 2, NL-7990 AA Dwingeloo, NL\\
        $^{3}$ Kapteyn Instituut, University of Groningen, PO Box 800, NL-9700 AV Groningen, NL\\
        $^{4}$ Astronomy Department, University of Cape Town, Rondebosch 7700, Cape Town, South Africa\\
        $^{5}$ Department of Astronomy, Columbia University, Mail Code 5246, 550 West 120th Street, 10027, New York, NY, USA \\
        $^{6}$ Center for Computational Astrophysics, Flatiron Institute, 162 5th Ave, 10010, New York, NY, USA\\
        $^{7}$ School of Physics, University of the Witwatersrand, 2050, Johannesburg, South Africa\\
        $^{8}$ Astronomisches Institut der Ruhr-Universit\"{a}t Bochum, Universit\"{a}tsstr. 150, D-44780 Bochum, Germany \\
        $^{9}$ SKA South Africa, 3rd Floor, The Park, Park Road, Pinelands 7405, South Africa\\
        $^{10}$ Department of Physics and Electronics, Rhodes Centre for Radio Astronomy Techniques and Technologies, Rhodes University, PO Box 94, Grahamstown 6140, South Africa\\
        $^{11}$ Max-Planck Institut f\"{u}r Radioastronomie, Auf dem H\"{u}gel 69, D-53121 Bonn, Germany\\
        $^{12}$ Centre for Space Research, North-West University, Potchefstroom 2520, South Africa\\
        $^{13}$ School of Physics and Astronomy, Cardiff University, The Parade, Cardiff CF24 3AA, UK

E-mail: \email{pserra@oa-cagliari.inaf.it}}

\abstract{We present the science case and observations plan of the MeerKAT Fornax Survey, an \hi\ and radio continuum survey of the Fornax galaxy cluster to be carried out with the SKA precursor MeerKAT. Fornax is the second most massive cluster within 20 Mpc and the largest nearby cluster in the southern hemisphere. Its low X-ray luminosity makes it representative of the environment where most galaxies live and where substantial galaxy evolution takes place. Fornax's ongoing growth makes it an excellent laboratory for studying the assembly of clusters, the physics of gas accretion and stripping in galaxies falling in the cluster, and the connection between these processes and the neutral medium in the cosmic web.

We will observe a region of $\sim12$ deg$^2$ reaching a projected distance of $\sim1.5$ Mpc from the cluster centre. This will cover a wide range of environment density out to the outskirts of the cluster, where gas-rich in-falling groups are found. We will: study the \hi\ morphology of resolved galaxies down to a column density of a few times $10^{19}$ cm$^{-2}$ at a resolution of $\sim1$ kpc; measure the slope of the \hi\ mass function down to \mhi\ $\sim5\times10^5$ \msun; and attempt to detect \hi\ in the cosmic web reaching a column density of $\sim10^{18}$ cm$^{-2}$ at a resolution of $\sim10$ kpc.}

\FullConference{MeerKAT Science: On the Pathway to the SKA,\\
	25-27 May, 2016,\\
		Stellenbosch, South Africa}

\begin{document}

\section{Background and main goals of the MeerKAT Fornax Survey}

\noindent Galaxies form and evolve at favorable locations of the cosmic web -- a large-scale network of intersecting sheets and filaments of matter which originates from weak density fluctuations in the early Universe and evolves under the pull of gravity (e.g., \cite{huchra1983,springel2005}). The large range of densities where galaxies form and live in the cosmic web corresponds to a broad variety of physical conditions in their environment: the temperature and pressure of the surrounding inter-galactic gaseous medium; the number density of galaxies; and the motion of galaxies relative to one another and to the inter-galactic medium. These conditions have a strong effect on the flow of cold gas -- the raw material for star formation -- in and out of galaxies and, therefore, on their evolution.

The clearest manifestation of the importance of environment for galaxy evolution is that red, gas-poor early-type galaxies become more common, and blue, gas-rich late-type galaxies less common, with increasing environment density. This is the morphology-density relation originally discovered by \cite{dressler1980} and since then revisited by numerous authors (e.g., \cite{postman1984,cappellari2011b}). The morphology-density relation and its evolution across cosmic time (e.g., \cite{dressler1997,fasano2000}) indicate that, in dense environments, galaxies tend to quickly lose their cold gas, stop forming new stars and become red and smooth. On the contrary, in low-density environments, galaxies can hold on to their cold gas and/or keep accreting fresh gas. This gas is then converted into new stars, giving late-type galaxies their blue colour, clumpy appearance and -- since gas is typically distributed on a disc -- their flattened shape. Understanding the cause of the strong connection between individual galaxies and the cosmic web in which they are embedded -- the very structure of the Universe -- is at the core of modern astrophysics.

The goal of the MeerKAT Fornax Survey is to study the physical processes that drive the evolution of galaxies in the densest regions of the cosmic web, galaxy clusters. How do these galaxies lose their cold gas, and why do they stop accreting fresh gas? We will study these processes in action by observing the flow of cold gas inside a particular cluster of galaxies, Fornax. Fornax is an ideal target because of its proximity, size and evolutionary state, and is perfectly located for MeerKAT observations (Table \ref{tab:summary}).

We will use MeerKAT to observe the neutral hydrogen (\hi) gas in Fornax. This is the gas phase from which molecular gas and new stars form, and is therefore a key observable to understand galaxies. A crucial property of the \hi\ gas is that it can be detected all the way out to the outskirts of galaxies. These regions are the interface between galaxies and their environment, and offer a privileged view of the way cold gas flows in and out of them. If imaged with sufficiently high sensitivity and resolution, \hi\ can indeed reveal on-going episodes of gas removal as well as accretion. Such direct evidence is of paramount importance to understand how environment influences galaxy evolution.

The main goals of the MeerKAT Fornax Survey are to:

\begin{itemize}
\item Detect the tails of \hi\ gas removed from galaxies living in Fornax. We will study the detailed physics of gas removal from galaxies of different mass and type as a function of their position within this cluster, and quantify the impact of these gas stripping events on the subsequent evolution of these objects.
\item Study the \hi\ mass function in Fornax. We will measure how much gas is contained in small galaxies and is available to be captured by larger ones fuelling star formation, and we will compare our results to those obtained in lower- and higher density environments.
\item Detect the faint \hi\ gas which, according to cosmological simulations, should be found between galaxies in the cosmic web. This is the most challenging goal, potentially providing the first images and velocity fields of this gas component across hundreds of kpc.
\end{itemize}

\noindent These are all key science goals spelled out in the recently revised Square Kilometre Array science case (\cite{blyth2015,deblok2015,popping2015}). With the MeerKAT Fornax Survey, we will start pursuing these goals already in the early phase of MeerKAT's life. We will do so by mosaicking the Fornax region out to 1.5 Mpc from its centre (projected), reaching a column density sensitivity between $\sim10^{18}$ and $\sim5\times10^{19}$ cm$^{-2}$ at resolutions from $\sim10$ to $\sim1$ kpc, and with an \hi\ mass detection limit of $\sim5\times10^5$ \msun.

\begin{table}
\centering
\caption{Key properties of Fornax}
\begin{tabular}{rrr}
\hline
RA & 3h 38m &\\ 
Dec & $-35$d 30m &\\ 
distance & 20 Mpc & \cite{ferrarese2000}\\ 
scale & 10 arcsec/kpc &\\ 
$M_\mathrm{dyn}$ & $7\times10^{13}$ \msun& \cite{drinkwater2001b}\\ 
$L_\mathrm{X}$ & $5\times10^{41}$ erg/s & \cite{jones1997} \\ 
$\sigma$ & 370 km/s & \cite{drinkwater2001b} \\ 
$R_\mathrm{vir}$ & 0.7 Mpc & \cite{drinkwater2001b}\\ 
\hline
\end{tabular}
\label{tab:summary}
\end{table}

\section{The fate of cold gas in clusters}

\subsection{Hydrodynamical and tidal effects}

\noindent Using the new MeerKAT data we will study a wide variety of physical processes which can influence the flow of cold gas in and out of galaxies, and are the likely drivers of the morphology-density relation. First, as the environment density increases so does the number of galaxies per unit volume. This implies an enhanced chance of encounters and, therefore, interaction (or even merging) between them. During these events tidal forces can remove \hi\ as well as stars from the outer regions of galaxies, reducing their future ability to form stars. In extreme cases these forces can also trigger strong star formation activity, which can quickly exhaust the initial cold gas reservoir of the interacting systems.

Tidal stripping of cold gas is observed frequently in groups of galaxies (e.g., \cite{shostak1984,english2010}), and there are strong indications that it is at work also in the outskirts of big clusters like Virgo (\cite{chung2009}). However, this stripping mechanism is inefficient when galaxies move at high speed relative to one another. Therefore, it is expected to become less important towards the centre of clusters and in clusters of high mass.

\begin{figure}
\begin{center}
\includegraphics[width=8.3cm]{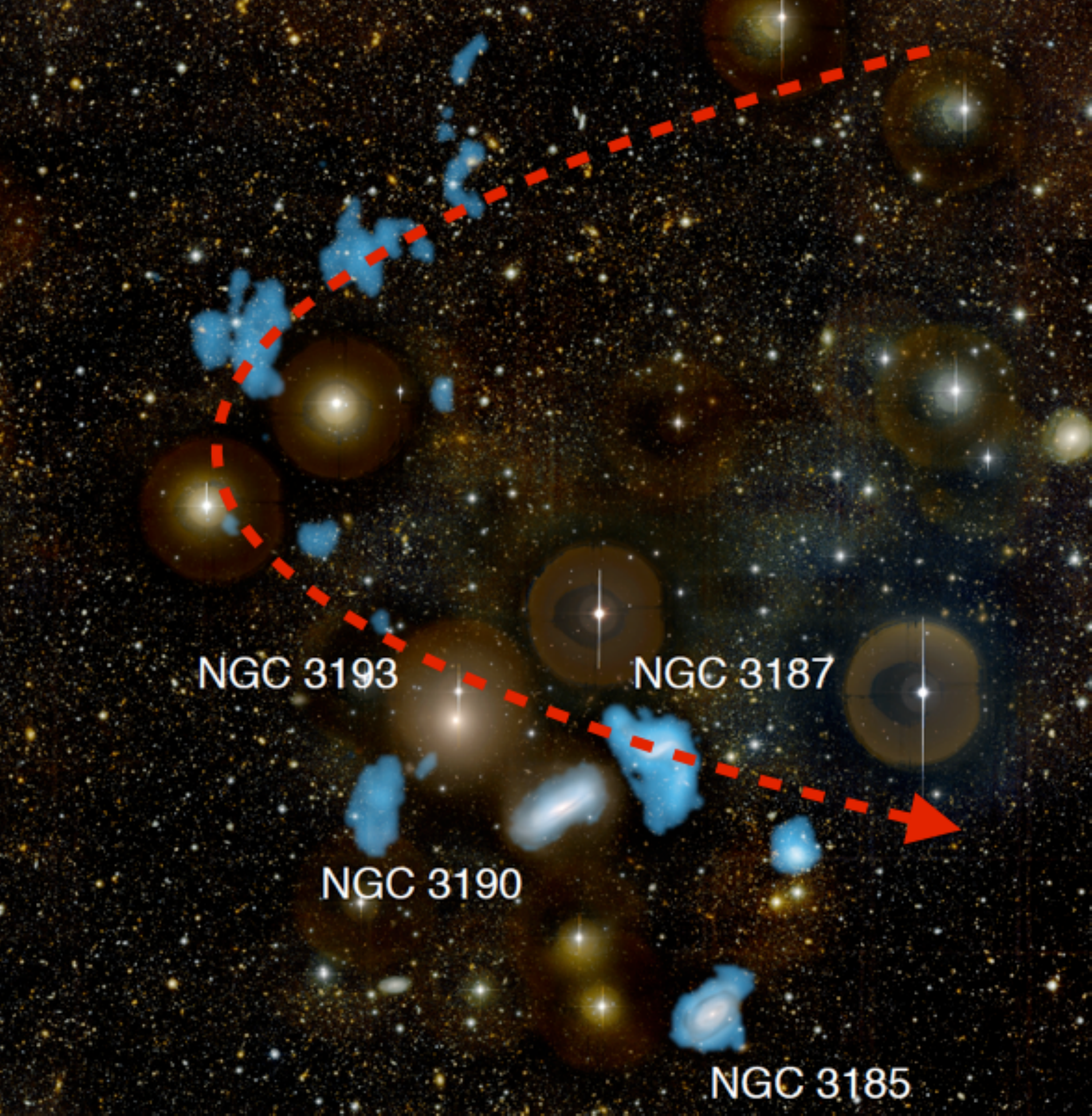}
\end{center}
\caption{\hi\ (light blue) on top of an optical image of HCG 44 (\cite{serra2013}). The orbit of the stripped galaxy is shown by the red line. The \hi\ column density is below $10^{20}$cm$^{-2}$. (Image credits: P.A. Duc.)}
\end{figure}

A second mechanism of \hi\ removal is related to the fact that the volume between galaxies is filled with a hot gaseous medium. In the densest regions of the cosmic web this medium is so hot that it can be detected in X rays (\cite{sarazin1988}). As galaxies move within a cluster the hot medium exerts a ram pressure on their gas but not on their stars, resulting in the stripping of gas alone (\cite{gunn1972}). \cite{kenney2004} show a particularly convincing example of this process in the Virgo cluster, NGC~4522, where the morphology of the \hi\ reveals with remarkable clarity the direction of motion of the galaxy relative to the hot medium. Such cases highlight the uniqueness of \hi\ observations for these studies.

Ram pressure stripping has so far been detected in massive clusters like Virgo (\cite{chung2009}), Coma (\cite{bravoalfaro2000}) and Abell 1367 (\cite{scott2010}). The strength of ram pressure grows linearly with the density $\rho$ of the hot medium and quadratically with the velocity $v$ with which galaxies move through it ($P_\mathrm{ram} \propto \rho \ v^2$). Both $\rho$ and $v$ grow with cluster mass and, therefore, ram-pressure stripping is expected to be efficient in big clusters. On the contrary, evidence of ram pressure has so far been weak in smaller clusters or groups of galaxies (e.g., \cite{westmeier2013}).

Galaxy tidal interaction and ram-pressure are the most frequently discussed mechanisms for the removal of \hi\ from galaxies. However, other processes are at work too. For example, the large-scale gravitational potential of groups or clusters of galaxies (or even the time-varying potential of merging groups or clusters) exerts tidal forces on galaxies that move through it. These forces can remove \hi\ and stars from the outskirts of galaxies, as proposed by \cite{bekki1999}. According to \cite{bekki2005a}, this may well be the origin of the famous Leo Ring, a giant ring of \hi\ discovered by \cite{schneider1983} around a small group of galaxies. Another system which may have formed through this mechanism is the giant \hi\ tail around the galaxy group HCG 44 (\cite{serra2013}, Fig. 1). Modelling suggests that this type of tidal stripping can be fast and efficient, removing large fractions of the \hi\ in $\sim1$ Gyr.

Additional stripping mechanisms are, again, related to the presence of hot gas between galaxies. For example, viscosity and turbulence can contribute to the stripping of \hi\ from a galaxy moving through the hot medium (\cite{nulsen1982}). Furthermore, the cold gas of galaxies could undergo thermal evaporation by interacting with the hotter, surrounding gas (\cite{cowie1977}). Indications of both types of stripping are found inside the Virgo cluster, where some galaxies are surrounded by regular but low-surface-density gas discs (\cite{cayatte1994,chung2009}). As with ram pressure, the expectation is that these mechanisms play a much bigger role inside massive clusters than in small clusters and groups.

In summary, the relative strength of the processes described above changes as a function of environment density: tidal forces and ram pressure should be the dominant effects in low- and high-density environments, respectively. However, observations suggest that there is significant overlap between the two, such as at the outskirts of massive clusters like Virgo (\cite{chung2009}) and Abel 1367 (\cite{scott2010}). In some cases the two types of stripping may even work together: gas could be moved to large radius by tidal forces and be susceptible to ram-pressure stripping because of its weaker gravitational binding to the galaxy.

Gas stripping from hydrodynamical and tidal forces, and its possible effect on star formation in galaxies, have received significant attention from theorist and simulators in recent years. Outstanding aspects that are being actively debated include: the efficiency of gas stripping of small galaxies relative to large galaxies in groups and clusters of different mass (e.g., \cite{bahe2015}); the role of magnetic fields in reducing the efficiency of ram-pressure stripping (\cite{shin2014,tonnesen2014}); the possible enhancement of star formation in stripped galaxies as well as within the tails of stripped gas (\cite{kapferer2009,tonnesen2012,roediger2014,henderson2016}); and how to use high-resolution simulations and observations to calibrate the recipes implemented in semi-analytic models of galaxy evolution (\cite{steinhauser2016}). Our MeerKAT survey, with its combination of sensitive spectral-line and radio continuum data (see below), will provide important observational constraints on these simulations.

\subsection{Removal of gas from small galaxies: the \hi\ mass function}

\noindent An interesting effect of the various gas stripping processes described above is that high-density environments may be devoid of small, gas-bearing galaxies. These are important because, when captured, they can progressively refill the gas reservoir of larger galaxies. If all small galaxies in clusters are \hi-poor, large galaxies will progressively lose all their \hi\ to, e.g., ram pressure or tidal forces without being able to accrete new gas -- hence ceasing forming stars.

The reason why small galaxies could be particularly gas-poor in clusters is that, having a shallow gravitational potential, they cannot hold on to their \hi\ even in the presence of weak stripping forces. An effective way to test this expectation is to compare the faint-end slope of the \hi\ mass function in clusters to that in lower-density environments. Results so far have been surprisingly contradictory. For example, \cite{zwaan2005} report an opposite effect (i.e., a steeper slope and a higher fraction of objects with a low \hi\ mass in denser environments), while \cite{springob2005} obtain results more in line with the expectations. Furthermore, \cite{moorman2014} and \cite{jones2016} find no significant difference between samples of galaxies living in different environments. Finally, the careful analysis of individual over-densities in the nearby large-scale structure using sensitive data, as done for the Ursa Major cluster (\cite{verheijen2001}) and for a sample of local groups (\cite{pisano2011}), suggests that the mass function is flat and low-\mhi\ galaxies are rare in at least some dense environments. Our data will establish whether that is the case in Fornax too.

\subsection{\hi\ accretion from the cosmic web}

\noindent According to modern cosmology, clusters like Fornax are situated at the intersection between filaments of the cosmic web. These filaments are filled with hydrogen gas, which is expected to stream towards galaxies and feed their continuous growth through star formation. Such accretion of gas from the cosmic web is a fundamental and yet highly controversial aspect of modern cosmology. On the one hand, it seems to be required to explain the rate of galaxy growth across cosmic time. On the other hand, attempts to image this  gas in the nearby Universe have so far been unsuccessful. This is partly due the fact that the density of gas in the cosmic web is too low for self-shielding against ionising radiation, and most of the gas should be ionised (\cite{rauch1998}). Although neutral gas has been detected in absorption along a limited number of QSO sight-lines (\cite{tumlinson2013}) its full distribution and kinematics -- only accessible through \hi\ in emission -- have never been studied before (\cite{popping2015} and references therein).

The detection of \hi\ in the cosmic web is extremely challenging because this gas is expected to have very low column density, $\sim10^{18}$ cm$^{-2}$ and below ($>1000$ times less dense than the \hi\ found inside star-forming galaxies). In contrast, the typical sensitivity of current interferometric \hi\ images is of a few times $10^{19}$ cm$^{-2}$. Single-dish radio telescopes can reach lower formal column densities but the resolution is so poor (many tens of kpc) that the \hi\ signal from the cosmic web can be diluted (\cite{popping2015}). With a sensitivity of $\sim10^{18}$ cm$^{-2}$ at the ideal resolution of $\sim10$ kpc over a survey area of 2 Mpc$^2$, our survey of the Fornax cluster is well positioned to attempt the detection and imaging of this elusive gas component in a cluster.

\section{The importance of Fornax}

\noindent Fornax is an extremely important system for the study of galaxy evolution in dense environments. First, it is very nearby -- the second most massive galaxy concentration within 20 Mpc after Virgo (Table 1). This means that the processes driving galaxy evolution in dense environments summarised in Sec. 2 can be studied at a unique resolution and sensitivity. Second, Fornax is in a very active evolutionary state. This fact is sometimes overlooked in the literature because most large galaxies in Fornax are of early morphological type -- indicating that the cluster as a whole is already quite evolved. Yet, observations demonstrate that Fornax is currently growing by accretion of new gas-rich galaxies and galaxy groups, and there is evidence that many dwarf galaxies have just joined the cluster (e.g., \cite{drinkwater2001a,waugh2002}). \cite{scharf2005} even suggest that the cluster itself and a nearby galaxy group (centred on NGC 1316) are falling towards each other along a filament of the cosmic web, making this an extremely exciting region to study. Our MeerKAT survey will cover the full region involved in this process of cluster growth -- including the cluster centre and the NGC 1316 group; see Sec. 5 -- allowing us to catch the effect of cluster assembly on galaxy evolution in action.

The other crucial point is that Fornax represents an important class of dense, low-mass clusters where many galaxies live but which are poorly studied in \hi. Indeed, while a few dozen clusters have been observed at limited sensitivity and poor resolution in \hi, most of what we know about the detailed physics of cold gas stripping in high-density environments comes from the observation of just a handful of clusters -- all much more massive than Fornax. The most notable example is the very nearby Virgo (e.g., \cite{giovanelli1983,chung2009}; see Fig. 1). Additional examples are some even more massive clusters such as Coma and Abell 1367, whose study is however hampered by the ten times larger distance (\cite{bravoalfaro2000,scott2010}). Only $\sim20$ percent of all galaxies in the Universe live in clusters this large. All other galaxies live in less dense regions of the cosmic web and evolve in very different conditions. In particular, substantial exhaustion of galaxies' cold gas reservoir, star formation quenching and morphological transformation appear to occur in dark-matter halos of mass between $\sim10^{13}$ and $10^{14}$ \msun\ (e.g., \cite{lagos2011,catinella2013,blyth2015,rafieferantsoa2015,odekon2016}). The Fornax cluster is the best target within this mass range, while the more massive clusters studied so far (e.g., Virgo, Coma) are well above it.

In the less massive Fornax cluster, the density of the intra-cluster medium and the velocity dispersion of galaxies are 2 times lower than in Virgo, while the number density of galaxies is 2 times larger (\cite{jordan2007}). Therefore, low-velocity galaxy encounters are much more frequent, and the importance of tidal interactions relative to ram-pressure should be significantly increased. Fornax offers therefore the unique possibility to extend the baseline of clusters for which we have a detailed observational \hi\ picture, and make a crucial contrast to massive clusters. The goal of our project is to perform exactly this comparison between Fornax and more massive clusters.

A final, interesting aspect of Fornax is that it will allow us to study the cold gas content of the many elliptical and lenticular galaxies in the cluster. Although typically gas-poorer than spirals, these galaxies do contain some cold gas both within and outside clusters (\cite{young2011,serra2012a}) and the comparison between gas and stellar morphology and kinematics reveals a clear pattern of gas accretion and removal in these objects (\cite{davis2011b,serra2014}). Our sensitive data will result in the \hi\ detection of numerous such galaxies in Fornax.

\section{Radio continuum and polarisation}

\noindent Our data will provide excellent sensitivity to radio continuum emission from galaxies in Fornax. Star forming galaxies give rise to radio synchrotron emission through the production of relativistic electrons in supernovae. This emission correlates tightly with the far-infrared emission, connecting the heating of dust by massive stars to the supernovae rate. As cluster galaxies are stripped of their inter-stellar medium this correlation is affected locally because of a lower radio continuum emission at the location of the strongest ram pressure (\cite{murphy2009}). Using our sensitive MeerKAT data we will be able to investigate this phenomenon further. Moreover, we will be able to study the possible enhancement of star-formation rate due to ram pressure.

\begin{figure*}
\begin{center}
\includegraphics[width=11cm]{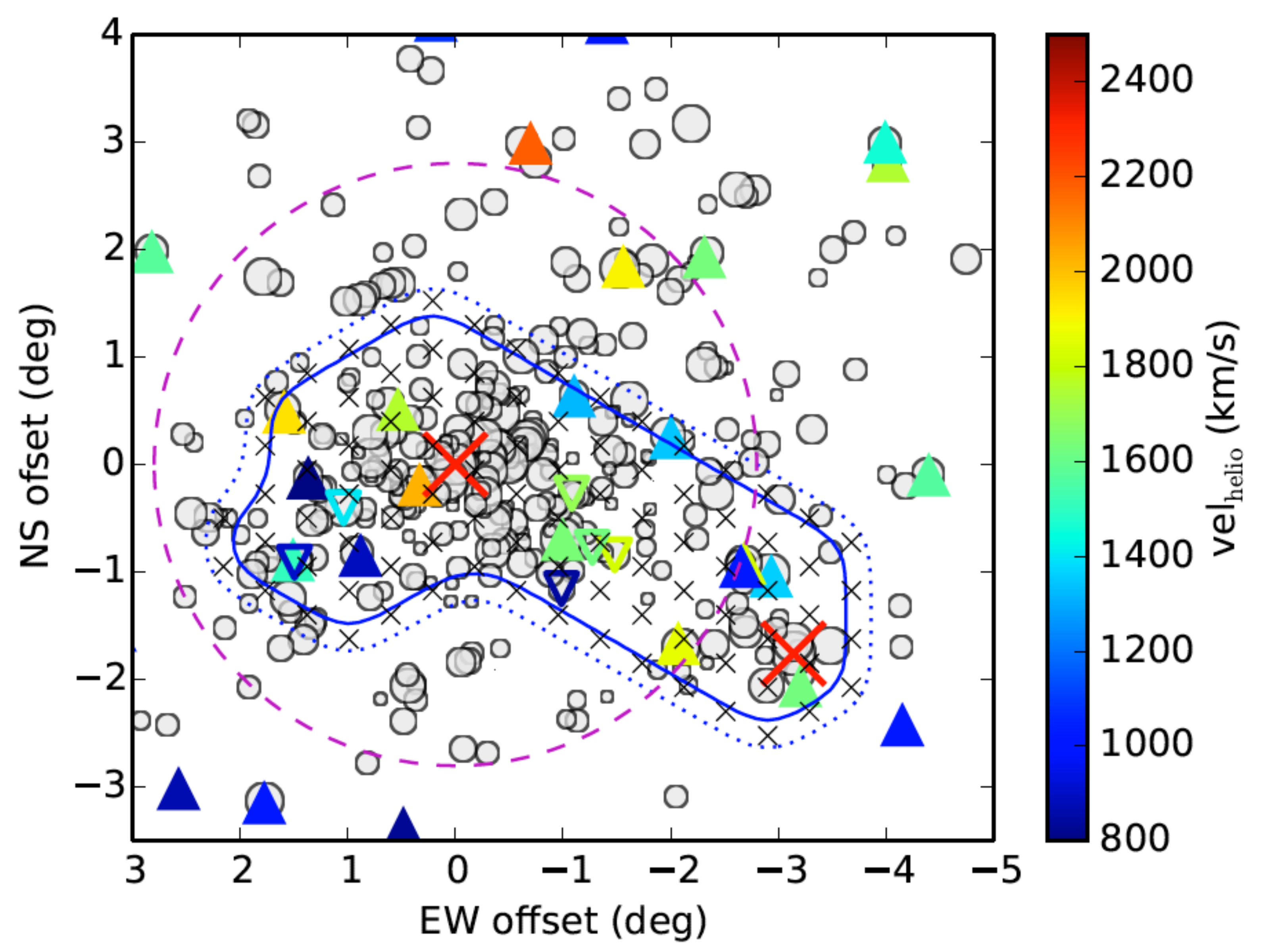}
\caption{MeerKAT Fornax Survey layout. North is up, east is left. The small black crosses represent 86 individual MeerKAT pointings.  The solid and dotted blue contours represent noise levels 10 percent and 40 percent higher than the minimum noise in the mosaic (0.1 mJy/beam in a 5 km/s channel with natural weighting). The solid blue contour defines a mosaic area of 11.8 deg$^2$. The grey circles indicate galaxies listed as definite or likely cluster members in \cite{ferguson1989}. The size of the circles is proportional to the logarithm of galaxies' B-band flux. The two large red crosses indicate the location of the cluster centre (NGC~1399 to the east) and of the south-west sub-group (NGC~1316, to the west). Solid, up-pointing triangles indicate galaxies detected in \hi\ by \cite{waugh2002} down to  $\sim3\times10^8$ \msun. Open, down-pointings triangles indicate new \hi\ detections obtained with the ATCA down to  $\sim5\times10^7$ \msun\ (P.I. Serra). Colours represent the heliocentric velocity of the \hi\ detections. The dashed magenta circle has a radius of $\sim1$ Mpc, which is $\sim1.5\times$ the virial radius (Table 1).}
\end{center}
\end{figure*}

Deep continuum data will also provide an ideal dataset to study AGNs in Fornax. The link between radio AGNs and the environment in which they are embedded is very tight and is particularly important in clusters, where radio galaxies and the X-ray-emitting intracluster medium influence each other profoundly (e.g. \cite{murgia2011}). The intracluster medium regulates the expansion of radio galaxies and affects their morphology by confining and distorting the radio lobes. At the same time, the radio lobes may inflate cavities in the intracluster medium. The most notable radio continuum source in our field is Fornax A, which resides in the south-west infalling sub-group. Fornax A is well known for its large radio lobes and was recently discovered to be injecting cosmic rays in the cluster (\cite{fermilat2016}). Deep \hi\ and radio continuum imaging will provide insights into the effects of Fornax A on the surrounding cold gas, further investigating the mix of radio continuum, X-ray emission and \hi\  studied by \cite{horellou2001}.

Radio continuum emission not associated with specific galaxies will also be of great interest too. Indeed, synchrotron cluster radio relics represent the most spectacular example of cluster radio emission and are directly connected to the dynamics of galaxy clusters. These optically unidentified diffuse and extended radio sources are detected at the cluster peripheries, where they trace shock waves deriving from the infall of gas and merging of sub-clusters (\cite{ensslin1998,pfrommer2007}). The current knowledge about cluster radio relics is still limited and only a few examples are reported in the literature (\cite{feretti2012}). The high sensitivity achieved by our observations of Fornax will have the potential to image (and study the polarisation properties of) relic sources in this dynamic, low-mass cluster.

Finally, regardless of the detection of diffuse radio continuum emission from within the cluster, we will be able to measure the Faraday rotation caused by the Fornax' magnetic field for a few thousands background polarised radio sources -- resulting in a characterisation of the the magnetic field in the cluster with a resolution of a few tens of kpc. This is adequate to start connecting for the first time the properties of galaxies and the intra-cluster medium in Fornax with those of the magnetic fields in which they are embedded.

\section{Survey strategy}

\noindent In order to reach the science goals summarised in Sec. 1 we will use MeerKAT to mosaic a region of $\sim12$ deg$^2$ down to a natural r.m.s. noise level of 0.1 mJy/beam in a 5 km/s ($\sim25$ kHz) channel. This will deliver the required \hi\ column density and mass sensitivity to reveal episodes of gas stripping, study the \hi\ mass function and attempt the first detection and imaging of \hi\ in the cosmic web. Assuming $T_\mathrm{sys}/$\it efficiency \rm $=22$ K for the 64 13.5-m MeerKAT dishes, this requires $\sim900$ h of telescope time including a $\sim10$ percent calibration overhead.

The survey footprint is shown in Fig. 2 and includes the cluster central $\sim0.5$ Mpc (radius), plus a small extension out to $\sim0.7$ Mpc towards a gas-rich group to the south-east, and a larger extension out to $\sim1.5$ Mpc towards the infalling south-west group dominated by the peculiar galaxy NGC~1316. The velocity range covered by our data allow us to reach significantly larger distances from the cluster along the line of sight (see below).

Fig. 3 shows the column density sensitivity within the survey area as a function of angular resolution. This was obtained by performing a simulation with the CASA package, where we adopted the most recent MeerKAT specs and the planned $uv$ coverage per mosaic pointing. As required by our science goals, our MeerKAT data will be sensitive to an \hi\ column density of $\sim5\times10^{19}$ cm$^{-2}$ at a resolution of $\sim10$ arcsec ($\sim1$ kpc), allowing us to detect and study the detailed morphology of gas in galaxies. Furthermore, at a resolution of $\sim30$ arcsec ($\sim3$ kpc) we will have excellent sensitivity and will be able to detect the faintest \hi\ tails formed in past interactions and gas stripping events. Finally, we will start to explore the column density regime expected for gas in the cosmic web ($10^{18}$ cm$^{-2}$ and below) at the appropriate resolution of $\sim100$ arcsec ($\sim10$ kpc).

\begin{figure*}
\begin{center}
\includegraphics[width=8cm]{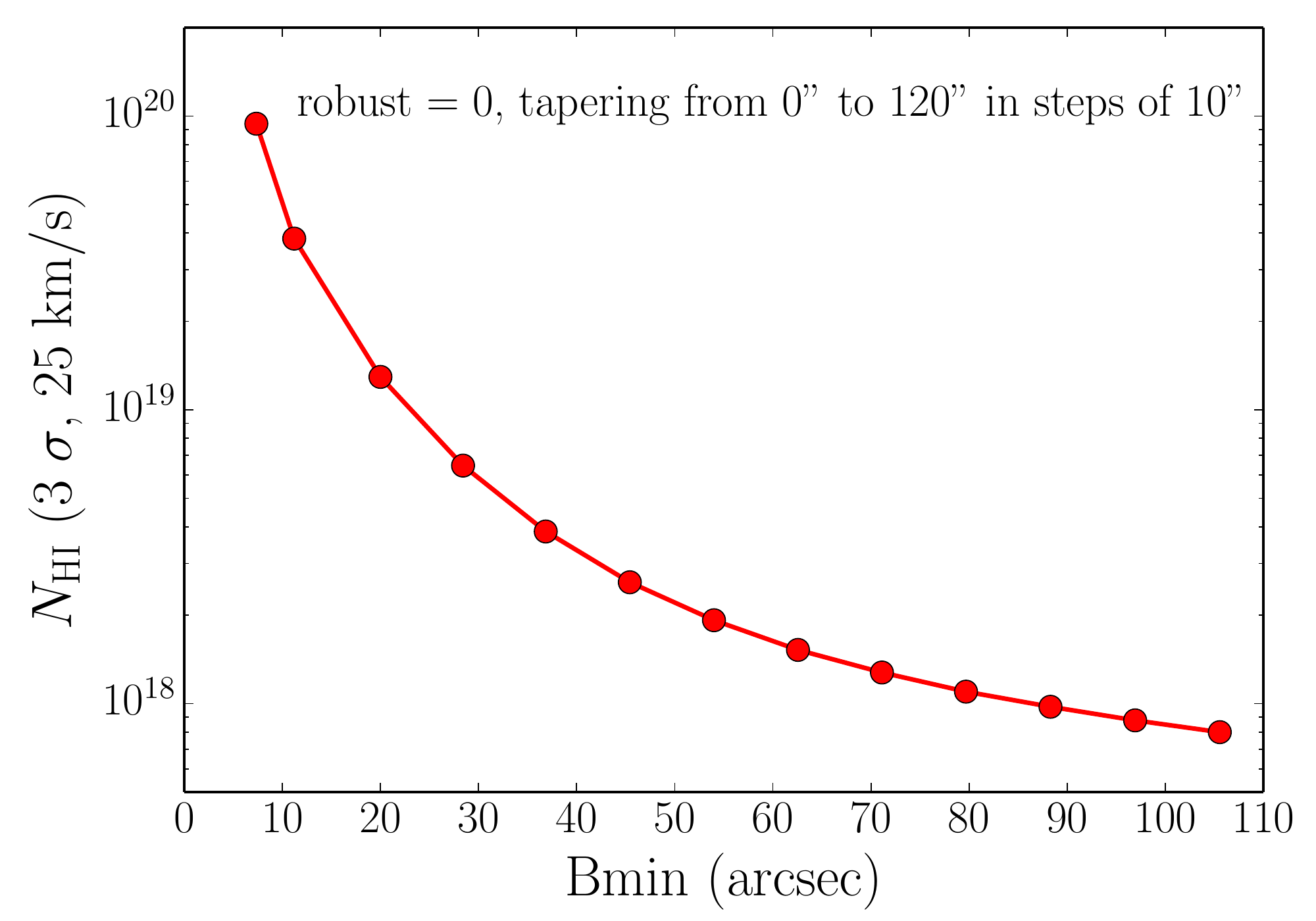}
\caption{Expected column density sensitivity as a function of angular resolution. Different angular resolutions are obtained tapering the visibilities at fixed Briggs weighting (robust = 0). The simulation was performed in CASA. At the distance of Fornax, 10 arcsec  $\sim1$ kpc (Table 1).}
\end{center}
\end{figure*}

With optimal angular resolution, the $3\sigma$ \mhi\ sensitivity of our survey within a 25-km/s line width (appropriate for dwarfs with a total baryonic mass of $\sim10^7$ \msun; e.g., \cite{lelli2016}) is $\sim5\times10^5$ \msun. Assuming that the \hi\ mass function of Fornax is the same as that in the field down to this sensitivity (i.e., $slope = -1.3$; \cite{martin2010}), and based on the number of known massive \hi\ detections from the Parkes and ATCA telescopes in the survey area (twenty; see Fig. 2), we predict a total of $\sim150$ \hi\ detections. In case of a flat mass function ($slope = -1$) this number goes down to $\sim60$. Of course, measuring the number of detections and the slope of the \hi\ mass function in Fornax is precisely one of the key goals of our survey. In this respect, these predictions serve the sole purpose of showing that we will be sensitive to variations of the mass function slope within the range of values discussed in the literature. Based on the above numbers, the error on the mass function slope will be well below 10 percent.

In order to study the \hi\ kinematics of the smallest \hi\ detections -- whose rotational velocity can be as low as $\sim20$ \kms\ (and even lower in projection) our spectral-line data will have an \hi\ velocity resolution of $\sim1$ km/s. We will obtain data at this resolution across an \hi\ recessional velocity range of, at least, $-1000$ to $+3500$ \kms, covering the full range of galaxies in the cluster. These data are within the RFI-protected frequency range for \hi\ at $z = 0$.

To study the radio continuum emission (both in total and polarised intensity) and the magnetic fields in the Fornax cluster we will take data over a 770-MHz-wide band in the frequency range 900 to 1670 MHz. The 25 kHz resolution of these data will facilitate RFI flagging and allow the commensal use of our data (e.g., to study \hi\ in the background of Fornax). Our MeerKAT mosaic of the Fornax region will be Nyquist sampled at the highest observing frequency of 1670 MHz.

\section{Multi-wavelength observations of Fornax}

\noindent Our MeerKAT data represent a significant step forward for the study of \hi\ and radio continuum emission in the Fornax cluster. Compared to previous Parkes observations (\cite{waugh2002}) we will go $\sim500\times$ deeper in \mhi\ and we will increase the linear resolution by almost two orders of magnitude. More recent \hi\ observations carried out with the Australia Telescope Compact Array (P.I. Serra) had already improved the situation, allowing us to make make six new \hi\ detections compared to \cite{waugh2002} (Fig. 2). Importantly, these new detections were found in the proximity of known \hi-rich galaxies, indicating that there may be regions within the cluster where there is a larger availability of cold gas. Those same regions may therefore harbour many more \hi\ detections at the $\sim100\times$ better sensitivity (and $\sim10\times$ better resolution) of our MeerKAT survey compared to the ATCA one. The other important result of the ATCA survey is the first detection of an \hi\ tail in Fornax (Lee-Waddell et al., submitted). The higher angular resolution of the MeerKAT data will be fundamental to uncover more such systems and understand their origin within Fornax.

The recent ATCA survey is just one component of a large set of multi-wavelength observations available for this cluster. The most recent efforts include: \it i) \rm the FDS survey, a deep optical imaging survey carried out with OmegaCAM on ESO VST (P.I.'s Iodice and Peletier; \cite{iodice2016}) covering our full MeerKAT Fornax mosaic (and more) down to a surface brightness sensitivity of $\sim28$ mag/arcsec$^2$ in $u$, $g$, $r$ and $i$ bands (and even deeper with moderate smoothing of the images); this survey provides a key dataset to look for the optical counterpart of low-surface-brightness \hi\ features (and vice versa), and to study the optical and \hi\ properties of the faintest galaxies in the cluster, including the newly discovered large population of ultra-diffuse galaxies (e.g., \cite{munoz2015}); \it ii) \rm a Herschel survey of Fornax, which covers the central 16 deg$^2$ of the cluster and resulted in the detection of cold dust in 30 galaxies of both early- and late morphological type (\cite{davies2013,fuller2014}); \it iii) \rm several observations of the CO (J=1-0) molecular gas line in galaxies within Fornax using the Mopra single-dish (P.I. Smith) and ALMA (P.I. Davis); together with our MeerKAT survey and the Herschel data, these observations will build a complete census of the cool interstellar medium (dust, molecular gas and atomic gas) in Fornax; \it iv) \rm integral-field spectroscopy of dwarf and giant galaxies in Fornax with a variety of instruments, including WiFeS (\cite{scott2014}), SAMI (P.I. Scott) and MUSE (P.I. Sarzi).

In addition to these new projects, several older campaigns resulted in valuable datasets which will complement our radio data. For example: i) the spectroscopic survey of \cite{drinkwater2001b}, which could be expanded to targets selected from our new VST images; ii) Chandra X-ray imaging of the cluster core (\cite{scharf2005}); iii) archival UV GALEX imaging (\cite{martin2005}); iv) archival WISE images (\cite{wright2010}); and v) HST/ACS imaging of the 40 brightest galaxies in Fornax (\cite{jordan2007}). In addition, the ASKAP \hi\ survey WALLABY (\cite{koribalski2012}) will provide complementary data on the larger-scale distribution of \hi\ in the Fornax region down to \mhi\ $\sim3\times10^7$ \msun\ at a 30 arcsec resolution.

\section{Summary}

\noindent Decades of research on galaxies have formed a tentative picture of the role of environment in shaping their gas content. We expect galaxies to become progressively \hi-poorer and transition from a late- to an early-type morphology through a varying balance of processes as a function of position in the cosmic web. The history of \hi\ observations in galaxy clusters demonstrates the importance of combining good column density sensitivity and resolution to study these processes. Initial low-resolution observations led to the conclusion that the \hi\ mass of spirals becomes progressively lower as they approach a cluster's centre (e.g., \cite{chamaraux1980}). However, it was only thanks to higher-resolution images that such decrease of the \hi\ mass could be linked to the shrinking of the \hi\ discs (e.g., \cite{cayatte1990}). Finally, by pushing the sensitivity to low column density, we have been able to directly see the faint gas being removed from the disc outskirts, i.e., witness the very cause of the reduced \hi\ disc size and mass (\cite{kenney2004}).

The lesson of these developments is that, if imaged at sufficiently high resolution ($\sim1$ kpc) and sensitivity (a few times $10^{19}$ cm$^{-2}$), \hi\ represents a unique, direct tracer of these processes in action because it can reveal tails of stripped gas and truncated gas discs, while by reaching \mhi\ detection limits of $10^6$ \msun\ and below we can quantify variations in the mass-function slope. This is the motivation for a large number of completed and planned deep \hi\ surveys, which target from low-density regions of the cosmic web (e.g., \cite{kreckel2011}) to small groups (e.g., \cite{verdesmontenegro2001}), rich groups like Ursa Major (\cite{verheijen2001}), small clusters like Fornax (this project), and clusters of larger size like Virgo and Coma (e.g., \cite{chung2009,bravoalfaro2000}). Comparing and contrasting the \hi\ properties of galaxies living in such different environments is key to obtaining a coherent picture of galaxy evolution.

Within this context, the MeerKAT Fornax Survey will provide the most accurate view of the \hi\ and radio continuum properties of galaxies in the Fornax cluster. Our mosaic of $\sim12$ deg$^2$ will reach a projected distance of $\sim1.5$ Mpc from the cluster centre (and much larger, $\sim20$ Mpc, along the line of sight) and cover a wide range of environment density out to the outskirts of the cluster, where gas-rich in-falling groups are found. The sensitive MeerKAT data will allow us to: \it i) \rm study the \hi\ morphology of resolved galaxies down to a column density of a few times $\sim10^{19}$ cm$^{-2}$ at a resolution of $\sim1$ kpc, hunting for signs of gas stripping throughout the cluster and allowing a comparison to the distribution of other components of galaxies (e.g. stars, dust, molecular gas); \it ii) \rm detect large numbers of dwarfs and measure the slope of the \hi\ mass function down to \mhi\ $\sim5\times10^5$ \msun, which we will then compare to values found in different environments; and \it iii) \rm attempt to detect \hi\ in the cosmic web at a column density of $\sim10^{18}$ cm$^{-2}$ and a resolution of $\sim10$ kpc. In the future, observations of this type will become routine over much larger areas and over a larger range of environments with the Square Kilometre Array.

\

\noindent \small \bf Acknowledgments. \rm This project has received funding from the European Research Council (ERC) under the European Union's Horizon 2020 research and innovation programme (grant agreement No 679627). \normalsize

\bibliographystyle{JHEP}
\bibliography{../../myrefs.bib}

\end{document}